\documentclass[aps,pre,two column,amsmath,amssymb] {revtex4-2}
\usepackage{color,graphics}
\usepackage{graphicx}
\usepackage[sort&compress]{natbib}
\usepackage{hhline}
\usepackage{soul}

\begin{document}
\setcounter{totalnumber}{3}

\preprint{}

\title{Non-Gaussian Rotational Diffusion and Swing Motion of Dumbbell Probes in Two Dimensional Colloids}

\author{Jeongmin Kim$^{1,\dagger}$, Taejin Kwon$^{2}$, and Bong June Sung$^{3,*}$}
\affiliation{$^1$Department of Chemistry Education and Graduate Department of Chemical Materials, Pusan National University, Busan 46241, Republic of Korea \\ 
$^2$Department of Chemistry and Cosmetics, Jeju National University, Jeju 63243,
Republic of Korea \\
$^3$Department of Chemistry and Institute of Biological Interfaces, Sogang University, Seoul 04107, Republic of Korea}
\email{bjsung@sogang.ac.kr,$^\dagger$jeongmin@pusan.ac.kr}
\date{\today}

\begin{abstract}
Two dimensional (2D) colloids exhibit intriguing phase behaviors distinct from those in three dimensions, as well as dynamic heterogeneity reminiscent of glass-forming liquids. Here, using discontinuous molecular dynamics simulations, we investigate the reporting dynamics of dicolloidal dumbbell probes in 2D colloids across the liquid-hexatic phase transition, where hexagonal bond-orientational order (HBOO) extends to quasi-long-ranged one. The rotational dynamics of dumbbell probes faithfully capture the structural and dynamical features of the host: Brownian in the isotropic liquid, and non-Gaussian in the hexatic and solid phases, reflecting both HBOO and dynamic heterogeneity of the medium. In the 2D hexatic and solid phases, probe rotation reflects heterogeneity as the dumbbells sample multiple dynamical domains of the host system: in mobile domains, they undergo rotational jumps of $\pi/3$ in accordance with HBOO, whereas in immobile domains they librate within cages formed by surrounding discs. Such non-Gaussianity disappears upon reentrant melting of the host medium driven by size polydispersity, highlighting a close connection between HBOO and probe dynamics. Furthermore, probe dynamics reveal both coupling (at a single particle level) and decoupling (at an ensemble-averaged level) between translation and rotation: swing motion emerges as their primary diffusion mode, while the Debye-Stokes-Einstein relation breaks down regardless of how the rotational diffusion coefficient is defined.
\end{abstract}

\maketitle

\section{Introduction}
2D colloidal systems exhibit intriguing phase behavior and dynamics~\cite{RevModPhys.60.161,Zahn:2000p890,PhysRevLett.107.155704,Peng:2010p1493,Wang:2010p1583,JeongminPRL,Mishra2015,Yan2019,Kumar2023}. Unlike in three dimensions, 2D solids cannot possess a true long-range translational order because of long-range fluctuations, not following the Lindemann melting criterion~\cite{Lindemann,Mermin:1968p1610,RevModPhys.60.161,binder2005glassy}. An intermediate hexatic phase exists between the liquid and solid phases with a quasi-long-ranged hexagonal bond-orientational order (HBOO)~\cite{Kosterlitz:1973p1581, PhysRevLett.41.121, PhysRevB.19.2457, PhysRevB.19.1855}. Recent studies have revealed that the phase diagram is complex and nontrivial, even for monodisperse systems~\cite{Thorneywork2017,Downs2021,Tsiok2022,Williams2022,Gruber2025}. 

The HBOO that develops in the 2D hexatic and solid phases has been identified as being strongly correlated with the heterogeneous dynamics~\cite{vanderMeer2015}. In these phases, diffusion of 2D colloids is no longer Brownian with non-Gaussian displacement distribution $G_s(r,t)$~\cite{Zahn:2000p890,Shiba:2009p1550,PhysRevLett.92.035502}. Interestingly, such non-Gaussianity persists even in the diffusive regime, where $G_s(r,t)$ is presumed to return to a Gaussian distribution~\cite{JeongminPRL}. Translational diffusion in 2D colloids was thus concluded to be seemingly Fickian yet heterogeneous~\cite{JeongminPRL,JeongminJCP}, a phenomenon that has inspired recent theoretical frameworks, such as the concept of diffusing diffusivity~\cite{Chubynsky2014,Li2024,Wang2025}.

Such heterogeneous dynamics emerges in diverse systems, including dense colloidal suspensions, gels, porous materials, glasses, and even biological systems~\cite{EdigerAnnuRev,RevModPhys.83.587,WeeksRepProg,EdigerJPCReview,Manoharan2015,Weeks2016,Agosta2021,Zhang2023ma,Kang2025}. It has long been recognized as significant because it exhibits spatiotemporal fluctuations, often referred to as spatially heterogeneous dynamics~\cite{EdigerAnnuRev,RevModPhys.83.587,GranickPNAS,GranickNATMAT,Mishra2015,Dubey2020,Mandel2022,Dauchot2024,Tanaka2025}. The overall dynamics thus represents an average over several dynamic domains, some of which are mobile while others are immobile, with slow exchange between them. In such slowly varying heterogeneous environments, the timescales of molecular diffusion may depend on, or even be dominated by, environmental fluctuations, frequently resulting in unexpected non-Gaussian behavior in the diffusive regime~\cite{GranickPNAS,GranickNATMAT,JeongminPRL,SkaugPRL,GuanACSNano,HeACSNano,Rusciano2022,Ming2023,Kumar2023,Chen2024,McCuskey2025,Hyeyoung2025,Chen2025}. 

Experimental and computational studies have employed probes as reporters to investigate dynamic heterogeneity~\cite{Park2017,Molaei2018,Sun2019,Mazaheri2020,Rose2020,Mayer2021,Feng2021,Chun2021,Mandel2022,Zhang2023jacs,Fink2023}. The analysis of probe dynamics provides insights into the heterogeneous dynamics of host systems and can further elucidate molecular diffusion under strongly heterogeneous environments. For example, Edmond {\it et al.}~\cite{WeeksPNAS} examined the translational and rotational diffusion of tetrahedral colloidal probes in three-dimensional (3D) binary colloidal suspensions using particle-tracking techniques. In supercooled suspensions, the translational motion of the probes violated the Stokes–Einstein relation, depending only fractionally on viscosity, whereas the tetrahedral probes exhibited Brownian rotational motion, faithfully following the Debye-Stokes-Einstein relation (DSER). This decoupling between translation and rotation is one of the hallmarks of the dynamic heterogeneity, and the analysis of translational and rotational dynamics reported by the probes provide a useful link for understanding the host glassy dynamics.

The analysis of rotational dynamics reported by probes is, however, not always straightforward in the presence of dynamic heterogeneity. Chong {\it et al.}~\cite{ChongPRL} argued that defining the rotational diffusion coefficient $D_R$ for glassy systems is inadequate due to librational motions. In their simulation study of 3D binary dumbbell systems, tilted librations of rigid dumbbells were found to drive the mean-square angular displacement (MSAD) into a diffusive regime without genuine reorientation. This led to a discrepancy between the values of $D_R$ obtained from the Einstein and Debye formalisms, each highlighting different aspects of the dynamics~\cite{StillingerJCP,Kawasaki2019,Zendehroud2024}. Such ambiguity must be carefully considered when rotational dynamics violates the DSER~\cite{StillingerJCP,StarrPRE,ChongPRL,Mishra2015,Park2017,vonBlow2019,Kawasaki2019,Dubey2020,Mandel2022,Yong2023,Barbhuiya2023,Raj2024,Zhang2025,Zhang2025pccp}.

In this study, we investigate the reporting dynamics of dicolloidal dumbbell probes in 2D colloids across the liquid-hexatic phase transition. We find that the HBOO, together with slugguish and heterogeneous dynamics of the host colloids in the 2D hexatic and solid phases, gives rise to intriguing non-Gaussian rotational diffusion of the dumbbell probes. The dumbbells experience and report the dynamic heterogeneity of the host medium: in immobile regions, they librate within cages formed by surrounding discs, whereas in mobile regions they undergo intermittent rotational jumps of $\pi/3$ radians. The resulting distribution of angular displacements deviates strongly from the Gaussian form expected for Brownian rotors. Remarkably, these oscillations persist even when the mean-squared angular displacement grows linearly with time. Thus, the rotational dynamics of dumbbell probes is seemingly Fickian yet non-Gaussian, reflecting the structural and dynamical behavior of the host colloids. 

We also find that such dynamic heterogeneity of dumbbell probes leads to both coupling and decoupling between translation and rotation. As in systems with glassy dynamics, the DSER breaks down, indicating decoupling at an ensemble-averaged level. Notably, this breakdown is independent of the formalism used to compute the rotational diffusion coefficient, unlike in supercooled liquids~\cite{ChongPRL,WeeksPNAS,EdigerJPCReview}. At the single-molecule level, however, the dumbbells diffuse primarily via swing motion, in which translation and rotation are strongly coupled, in contrast to gliding motion (translation without rotation)~\cite{Oh2016}.

The rest of this paper is organized as follows: the model systems and simulation details are described in Sec.~\ref{sec:model}. Simulation results are presented and discussed in Sec.~\ref{sec:results}. Finally, summary and conclusions are presented in Sec.~\ref{sec:conclusion}.
  
\section{Models and Methods}\label{sec:model}
In this Section, we present the 2D hard-disc model system with dicolloidal dumbbell probes, the details of discontinuous molecular dynamics simulations, and the methods for computing dynamical properties.

\subsection{Models for dicolloidal probes and 2D colloids}
\begin {figure}
\centering\includegraphics [width=1\linewidth] {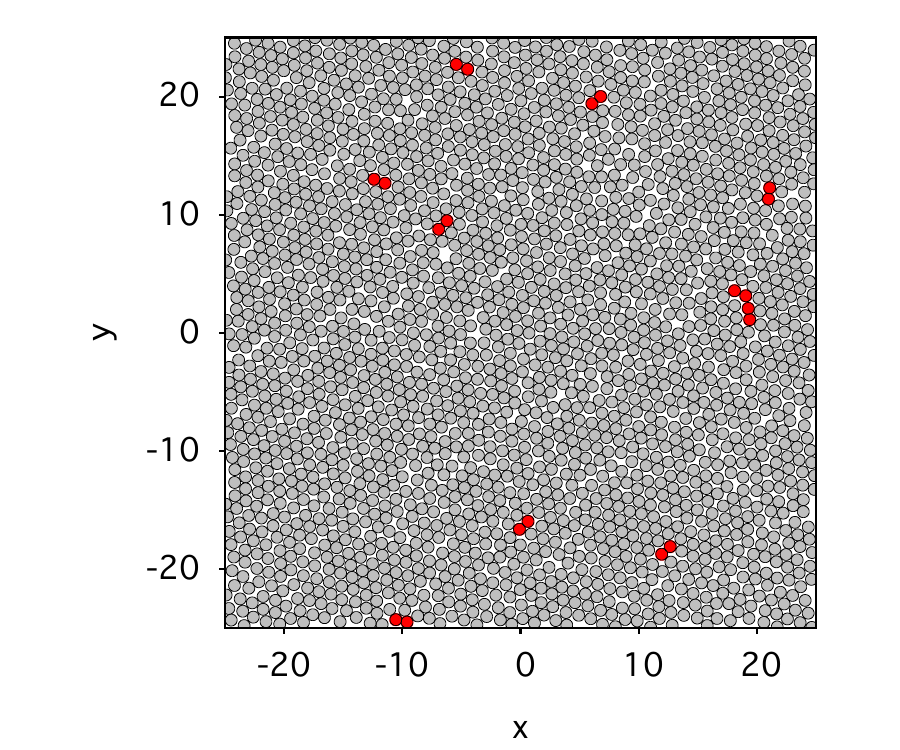}
\caption{A representative simulation snapshot of 2D monodisperse colloids (gray) and ten dumbbell probes (red).}
\label{fig:snap}
\end{figure}
We model a 2D colloid particle as a hard disc of a diameter $\sigma$ and mass $m$. A dicolloidal probe is modeled as a hard dumbbell: a dimer of hard discs of the same diameter and mass (Fig.~\ref{fig:snap}). The bond distance of a dumbbell is allowed to fluctuate from 0.95 to 1.05$\sigma$. In our simulations, $\sigma$ and $m$ are the units of length and mass, respectively. We examine 6 different area fractions ($\phi\equiv N \pi\sigma^2/4L^2$) to investigate all three phases in 2D from an isotropic liquid to a solid with $\phi\in[0.5,0.74)$, where $N$ is the number of discs and $L=50\sigma$ is the dimension of a simulation cell. We introduce ten probe dumbbells ($N_d = 10$), placing them in the system such that the area fraction ($\phi_d \equiv 2N_d \pi\sigma^2 / 4L^2$) of probes is only 0.006. $k_BT$ is the energy unit set to be unity with $k_B$ being the Boltzmann constant, and $T$ the temperature. The unit of time ($\tau$) is, therefore, $\tau \equiv \sqrt{m\sigma^2/k_BT} = 1$. In this paper, all quantities are expressed in reduced units.

According to the KTHNY theory based on the defect-mediated melting mechanism~\cite{Kosterlitz:1973p1581, PhysRevLett.41.121, PhysRevB.19.2457, PhysRevB.19.1855}, there are two continuous second-order transitions via the hexatic phase between the solid and the isotropic liquid . A melting transition occurs from the solid to the hexatic phase when a pair of dislocations dissociates. Then, a dislocation dissociates into disclinations when the hexatic transforms into the isotropic liquid phase. At $\phi$ = 0.7 and 0.72, hard discs experience the freezing transition between the isotropic liquids and the hexatic, and the melting transition between the hexatic and the solid, respectively. Bernard and Krauth reported that the liquid-hexatic transition is the first order transition with a very narrow single hexatic regime between $\phi$=0.716 and 0.720 for hard discs~\cite{PhysRevLett.107.155704}. The phase diagram has been shown to be complex and nontrivial, even in simple monodisperse systems~\cite{Thorneywork2017,Downs2021,Tsiok2022,Williams2022,Gruber2025}.

To investigate the effects of the HBOO on the rotational dynamics, we study polydisperse 2D colloidal systems, too. The size polydispersity ($\Delta$), in addition to packing density, controls the phase behavior of 2D colloids~\cite{SadrPRL,ItoPRE,Kawasaki:2011p1594,Tanaka:2011p1602}:
\begin{equation}
\Delta \equiv \frac{\sqrt{\overline{\sigma^2}-\bar{\sigma}^2}}{\bar{\sigma}},
\end{equation}
where $\overline{\cdots}$ denotes an average over particles. In this study, a colloid diameter lies between 0.5 and 1.5$\sigma$, and its distribution is Gaussian with the average diameter $\bar{\sigma} = 1\sigma$, and standard deviation $\Delta$. Then, $\phi \equiv N \pi\overline{\sigma^2}/4L^2=N\pi(\Delta^2+1)/4L^2$. In order to cover the melting transition, we consider three different values of $\Delta$ ($\approx$ 0.05, 0.09 and 0.13) at a fixed packing fraction of $\phi = 0.71$. It was reported that for Weeks-Chandler-Anderson colloids near $\Delta \approx 0.09$ in 2D, a dispersity-driven melting transition occured with a short-range BOO~\cite{SadrPRL,ItoPRE,Kawasaki:2011p1594}.

\subsection{Discontinuous molecular dynamics simulations}
Initial configurations are prepared by placing hard discs and dumbbells on a square lattice. We equilibrate and simulate the systems using discontinuous molecular dynamics (DMD) simulations~\cite{allen}. In the case of polydisperse systems, we carry out $NpT$ Monte Carlo simulations at high pressure until the desired $\phi$ is obtained. Polydisperse colloidal systems are then also equilibrated and evolved using DMD simulations. DMD simulation evolves the system via successive elastic collisions, employing an event-driven algorithm~\cite{allen,PhysRevE.69.051101,JeongminJCP}. The initial velocities of hard discs and dumbbells are chosen randomly according to the Maxwell-Boltzmann velocity distribution with zero total momentum, {\it i.e.}, the system is at rest. The initial configurations of each $\phi$ are equilibrated for at least 10 times the relaxation time. Periodic boundary conditions are applied in all directions.

\subsection{Computing dynamic properties: translation and rotation}
The translation of dumbbells and discs are studied using the mean-squared displacement ($\langle (\Delta r(t))^2 \rangle$) of the particles. $\Delta r(t)$ is the displacement of a particle during time $t$ and $\langle \cdots \rangle$ denote an ensemble average. For dumbbells, we track their centers of mass. $\langle (\Delta r(t))^2 \rangle$ is linear in time at long times. In the diffusive regime, the self-part of van Hove correlation function ($G_s(r,t) =  \langle \delta(r-|{\bf r}_i(t)-{\bf r}_i(0)| ) \rangle$) is expected to be Gaussian. Here, ${\bf r}_i(t)$ is the position vector of the $i$-th particle. However, for dense colloidal suspensions and supercooled liquids, $G_s(r,t)$ usually deviates from an expected Gaussian distribution even in diffusive regime.

To study the rotational diffusion of the dumbbell probes, we track an {\it unbound} orientational angle ($\varphi(t)$) of each dumbbell with time~\cite{StanleyPRL,WeeksPNAS}. We note that angles are in radian. The mean-squared angular displacement (MSAD, $\langle (\Delta \varphi (t))^2 \rangle$) is calculated using the relation $\langle (\Delta \varphi (t))^2 \rangle = \langle |\varphi(t) - \varphi(0)|^2 \rangle$. Its probability distribution ($G(\varphi,t)$) of the angular displacement is given by:
\begin{equation}\label{eq:gpt}
G(\varphi,t) = \langle \delta(\varphi-|{\bf \varphi}_j(t)-{\bf \varphi}_j(0)| ) \rangle,
\end{equation}
where $j$ is a dumbbell index. As for $G_s(r,t)$ in translation, $G(\varphi,t)$ is expected to be Gaussian in the rotationally diffusive regime (when $\langle (\Delta \varphi (t))^2 \rangle \sim t$) as follows:
\begin{equation}\label{GaussRot}
G(\varphi,t)=\frac{1}{\sqrt{4\pi D_Rt}}\exp \bigg(\frac{-\varphi^2}{4D_Rt} \bigg),
\end{equation}
where $D_R$ is a rotational diffusion constant from the Einstein formalism~\cite{liquids}. To quantify how much $G(\varphi,t)$ deviates from the expected Gaussian distribution~\cite{RahmanPR}, we calculate the rotational non-Gaussian parameter ($\alpha_{2,R}(t)$) of dumbbells as follows:
\begin{equation}\label{eq:nonG}
\alpha_{2,R}(t) = \frac{\langle (\Delta \varphi (t))^4 \rangle}{{3 \langle (\Delta \varphi (t))^{2}\rangle }^2} - 1.
\end{equation}
When $G(\varphi,t)$ is Gaussian, $\alpha_{2,R}(t) = 0$. 

We also estimate the time correlation function ($U(t) = \langle \vec{e}(t) \cdot \vec{e}(0) \rangle$) of the unit bod vector ($\vec{e}(t)$) of a dumbbell. According to the Debye approximation, $U(t)$ decays exponentially in the diffusive regime, \emph{i.e.}, $U(t) = \exp(-D_Rt)$ with a rotational diffusion coefficient, $D_R$~\cite{zwanzig}. We note that $U(t)$ is a reorientational correlation function with the first order Legendre polynomial. From the Gaussian approximation, $U(t)$ would also be $U(t) = \exp[-\langle(\Delta \varphi(t))^2\rangle/2]$. However, for dense liquids, $U(t)$ often becomes stretched, violating the Debye model. $U(t)$ then fits to the Kohlrausch-Williams-Watts (KWW) function:
\begin{equation}\label{eq:kww}
U(t)\approx\exp[{-(t/\tau_{KWW})}^\beta],
\end{equation} 
where $\beta$ is a stretching parameter of which value lies between 0 and 1~\cite{liquids,EdigerAnnuRev}.

\begin {figure}
\includegraphics [width=3.2in] {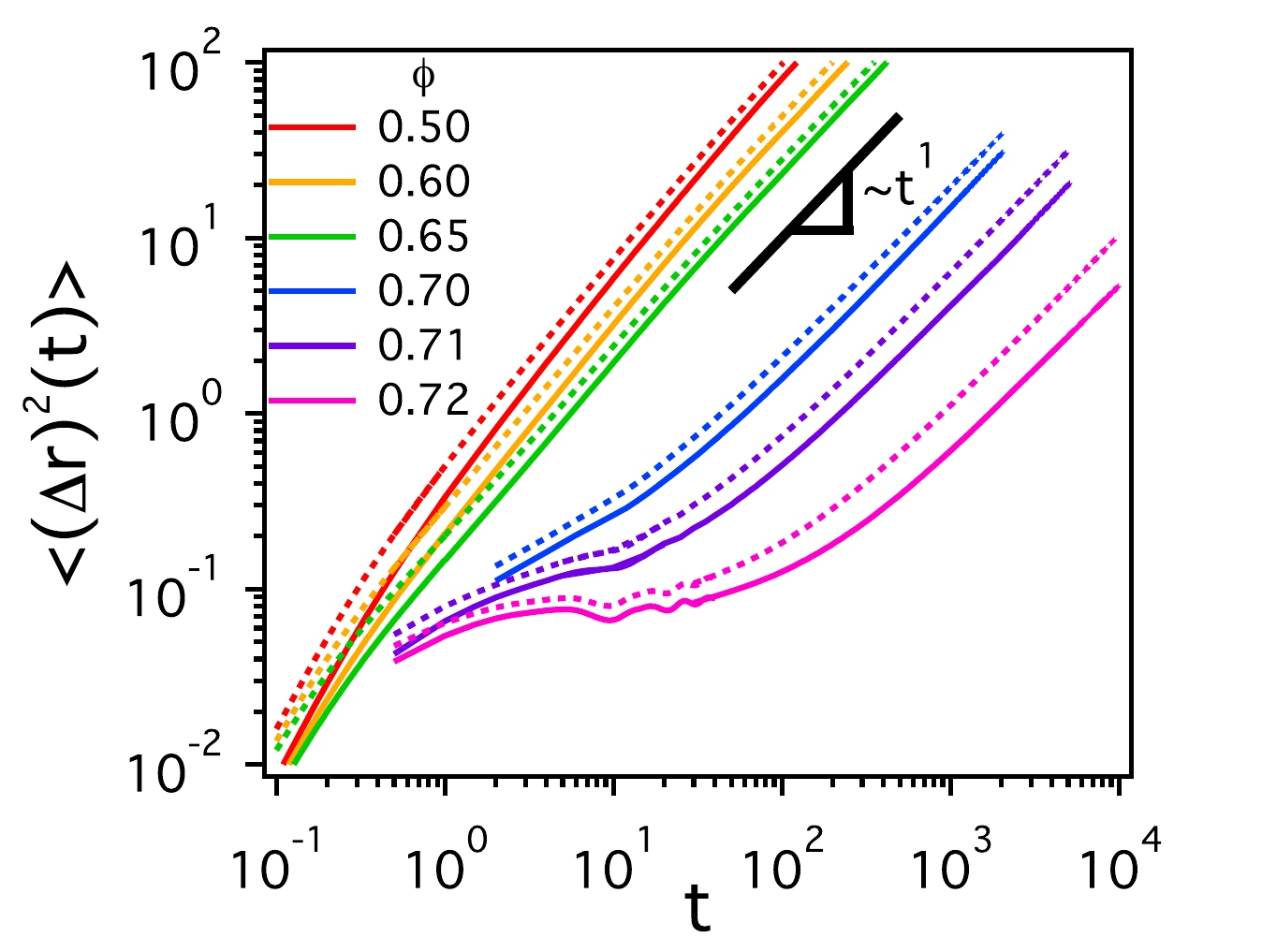}
\caption{Mean-squared displacements ($\langle (\Delta r(t))^2 \rangle$) of dumbbells (solid lines) and discs (dot lines) for different values of $\phi$.}
\label{msd}
\end{figure}

\section{Results and Discussions}\label{sec:results}
In this Section, we present the translational and rotational dynamics of dumbbell probes and their decoupling across the 2D freezing transition. In particular, we highlight the dynamic heterogeneity of the rotational motion, which manifests as intermittent jumps of $\pi/3$. We further identify swing motion as the primary diffusion mechanism of dumbbells in the presence of HBOO, resulting from translation–rotation coupling at the single-molecule level. Finally, we show that the reporting dynamics of dumbbells return to Brownian behavior during the 2D reentrant melting induced by size polydispersity.

\subsection{Dumbbell probes capture host dynamics and exhibit dynamic heterogeneity under HBOO}
Figure~\ref{msd} depicts the mean-squared displacements (MSDs, $\langle (\Delta r(t))^2 \rangle$) of dumbbell probes and discs at various $\phi$. It is clear that the MSDs of dumbbells closely follow those of discs at all investigated $\phi$, indicating that dumbbells are fairly good probes for examining the dynamics of discs in all three phases. The MSDs show distinct changes across the freezing transition at $\phi\approx0.7$. In contrast to the isotropic liquids, in the 2D hexatic and solid phases ($\phi > 0.7$), translation becomes dramatically slower, and $\langle (\Delta r(t))^2 \rangle$ exhibits a subdiffusive regime at intermediate time scales, {\it i.e.}, $\langle (\Delta r(t))^2 \rangle \sim t^b$ with time exponent $b < 1$. The subdiffusion is accompanied by spatial heterogeneity in dynamics, where particles with similar mobilities form dynamic clusters in localized regions~\cite{Zahn:2000p890,PhysRevLett.92.035502,JeongminPRL}. At $\phi = 0.72$, the MSD even displays a plateau over about two orders of magnitude in time. Such a plateau indicates that dumbbells and discs rattle in a transient cage formed by neighboring particles. At long times, the MSD in all cases eventually enters the diffusive regime with $b=1$ after the transient subdiffusion and rattling, in which one would presume that the intermediate scattering function fully decays and the non-Gaussian parameter also decays to zero. However, even in the diffusive regime, translational diffusion remains heterogeneous, with a non-Gaussian, oscillatory $G_s(r,t)$ observed in 2D hexatic and solid phases~\cite{JeongminPRL}.

\begin {figure}
\includegraphics [width=3.2in] {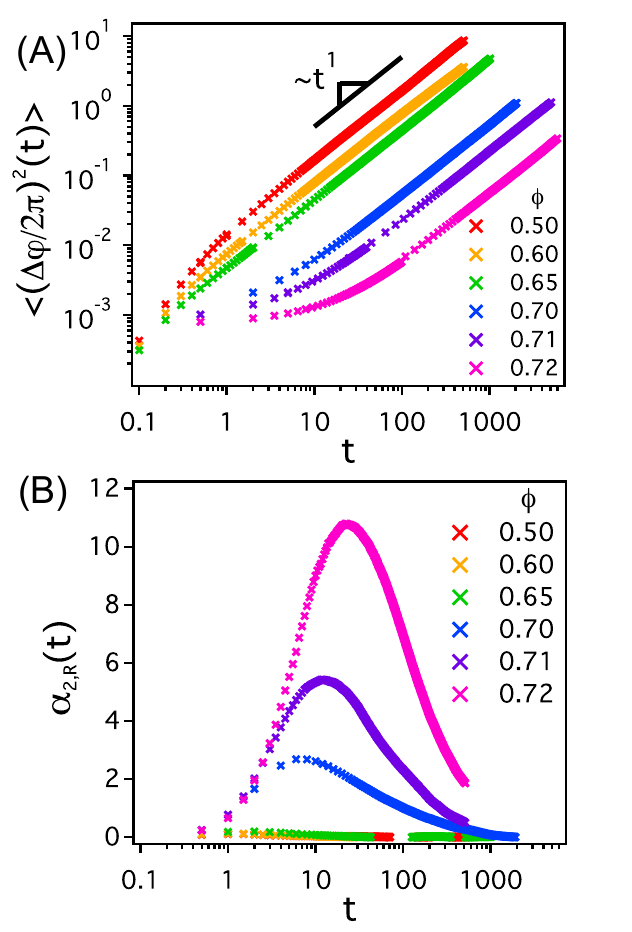}
\caption{Rotational dynamics of dumbbell probes. (A) Mean-squared angular displacement, $\langle(\Delta \varphi(t))^2\rangle$, divided by ${(2\pi)}^2$, and (B) a rotational non-Gaussian parameter, $\alpha_{2,R}(t)$ (Eq.~\ref{eq:nonG}).}
\label{fig:rotation}
\end{figure}

\begin {figure*}
\centering\includegraphics [width=6.0in] {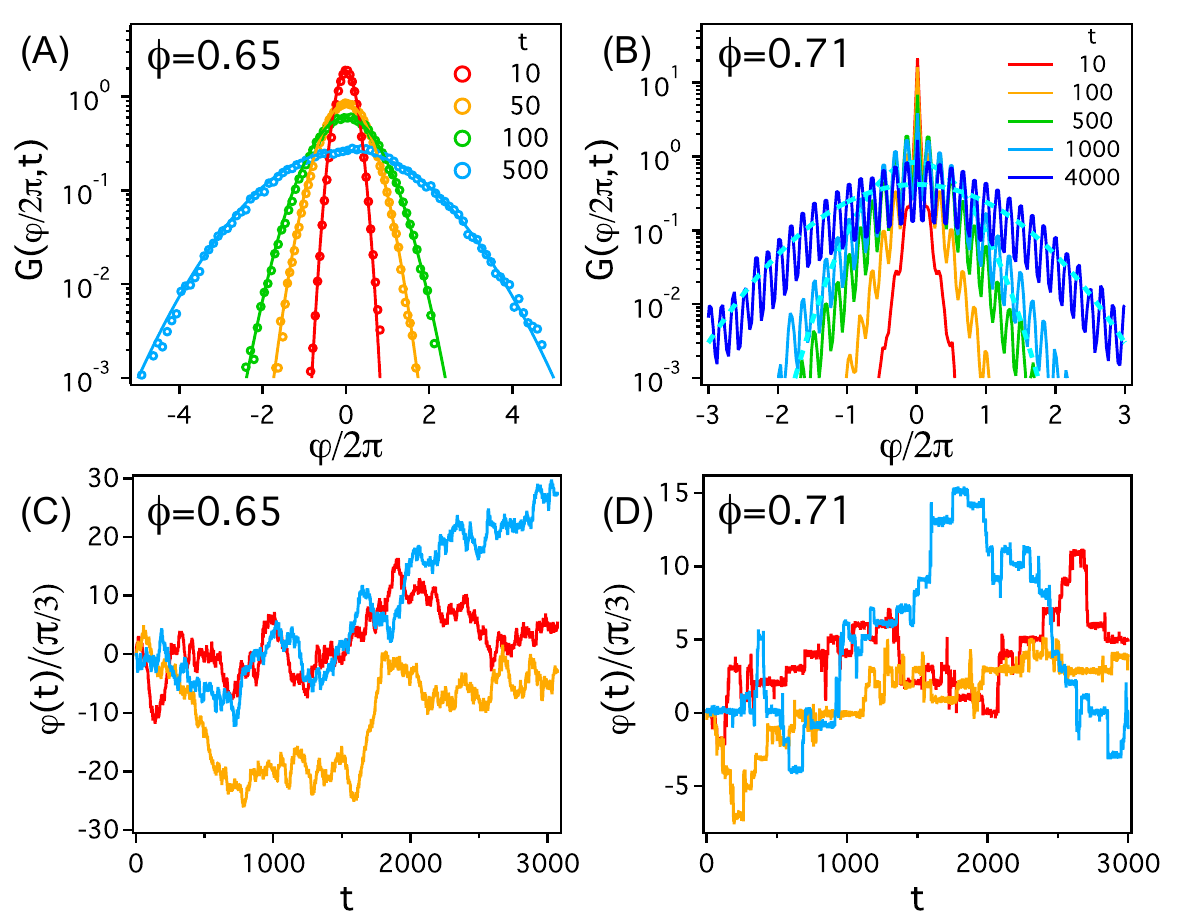}
\caption{Gaussian and non-Gaussian rotational dynamics of dumbbell probes across the 2D freezing transition. (A,B) Probability distributions $G(\varphi,t)$ (Eq.~\ref{eq:gpt}) of angular displacements of dumbbells at (A) $\phi = 0.65$ and (B) $\phi = 0.71$. Symbols denote simulation results; solid lines in (A) and dotted lines in (B) are Gaussian distributions (Eq.~\ref{GaussRot}) with the rotational diffusion constant obtained from the MSAD in Fig.~\ref{fig:rotation}(A). (C,D) Representative angular trajectories of dumbbells at (C) $\phi = 0.65$ and (D) $\phi = 0.71$. Here, $\phi(t)$ is the unbound rotational angle of each dumbbell.}
\label{fig:gstheta}
\end{figure*}

Figure~\ref{fig:rotation}(A) depicts the mean-squared angular displacements (MSADs, $\langle(\Delta\varphi(t))^2\rangle$) of dumbbell probes. Similar to the MSDs (Fig.~\ref{msd}), the MSADs also reflect the dynamic features of the discs in all three phases. After the freezing transition ($\phi \geq 0.7$), the MSADs exhibit a subdiffusive regime, and the rotation of the dumbbells slows down significantly. This subdiffusive rotation implies hindered motion by surrounding discs such that the dumbbells undergo rotational \textit{libration}, analogous to the rattling observed in translational motion. Accordingly, the rotational non-Gaussian parameter ($\alpha_{2,R}(t)$) becomes relatively large (Fig.~\ref{fig:rotation}(B)), indicating that the rotational displacement distribution function $G(\varphi,t)$ deviates significantly from a Gaussian distribution. At long times, the MSADs enter the diffusive regime, where $\alpha_{2,R}(t)$ decays back to zero, suggesting that $G(\varphi,t)$ becomes Gaussian at such timescales. As will be discussed in the next paragraph, however, $G(\varphi,t)$ remains highly oscillatory even at long times, rather than converging to a normal Gaussian.

As the HBOO strongly governs the dynamics of the host discs, $G(\varphi,t)$ can serve as a sensitive reporter of structural and dynamical changes in the host medium. We find that the rotational dynamics of the dumbbell probes captures well the dynamical changes across the freezing transition. In Fig.~\ref{fig:gstheta}, we compare $G(\varphi,t)$ of dumbbell probes before ($\phi=0.65$) and after ($\phi=0.71$) the freezing transition. At $\phi=0.65$, where the HBOO is short-ranged, $G(\varphi,t)$ is Gaussian at all times in the diffusive regime. Moreover, the computed $G(\varphi,t)$ matches well with the expected Gaussian distribution (Eq.~\ref{GaussRot}) using $D_R$ obtained from the MSAD. This indicates that dumbbell probes in the isotropic 2D liquids behave as Brownian rotors. At $\phi=0.71$, where the HBOO is quasi-long-ranged, however, $G(\varphi,t)$ is no longer Gaussian but highly oscillatory (Fig.~\ref{fig:gstheta}(B)). These oscillations persist even in the diffusive regime (e.g., $t = 4000$), where $\alpha_{2,R}(t) \approx 0$. This clearly indicates that the dumbbell probes are no longer Brownian rotors; their rotation appears Fickian but is in fact heterogeneous.

 \begin {figure*}
\centering\includegraphics [width=6in] {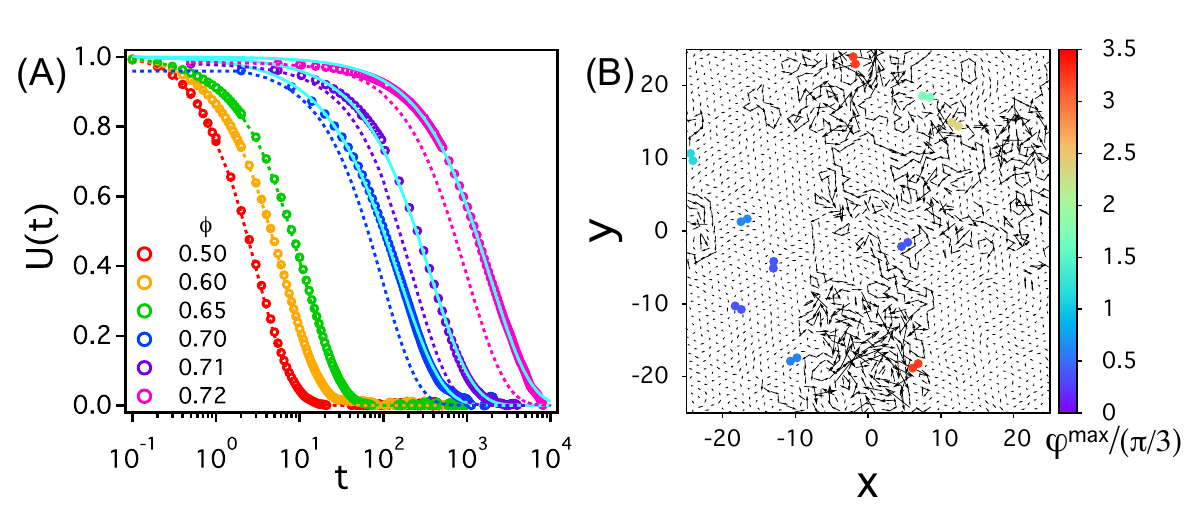}
\caption{Dynamic heterogeneity in the rotational dynamics of dumbbell probes. (A) Time correlation function $U(t)$ of the normalized bond vectors of dumbbells. Symbols denote simulation results, dotted lines indicate the Gaussian approximation \emph{i.e.}, $U(t) = \exp[-\langle (\Delta \varphi(t))^2 \rangle / 2]$, and cyan solid lines are fits to the stretched-exponential function (Eq.~\ref{eq:kww}) with $\beta = 0.85$, 0.88, and 0.90 for $\phi = 0.70$, 0.71, and 0.72, respectively. (B) Displacement map of 2D colloidal discs at $\phi = 0.71$ in the hexatic phase. Black vectors denote disc displacements during $\Delta t = 200\tau$, and filled colored circles represent dumbbell probes. The color of each dumbbell corresponds to $\varphi^{\mathrm{max}}/(\pi/3)$, its maximum rotational displacement normalized by $\pi/3$.}
\label{fig:tcf}
\end{figure*}

The angular trajectories $\varphi(t)$ in Figs.~\ref{fig:gstheta}(C) and~\ref{fig:gstheta}(D) highlight the clear difference in rotation of the dumbblls across the freezing transition. At $\phi = 0.65$, each $\varphi(t)$ represents a typical example of 1D random walk, whereas it exhibits intermittent jumps by $\pi/3$ at $\phi = 0.71$. Each trajectory shows long libration phases, associated with the plateau of the MSAD, in between rare and sudden jumps in $\varphi(t)$. As transient trapping by surrounding discs hinders the rotation, the dumbbells can rotate only when the neighboring discs escape from their cages. The magnitude ($\sim\pi/3$) of the rotational jump further reflects the local HBOO around the dumbbell in the hexatic phase. In fact, supercooled liquids often exhibit rotational jumps, but of about $\pi$ in the absence of strong structural order~\cite{StillingerJCP,StarrPRE,ChongPRL,LeporiniPRE,TakaePRE,HeejinPRE}. In the present study, the dumbbell probes can participate in forming the HBOO owing to their appropriate size. However, the reporting capability of the probes is not limited to such cases, as they also exhibit similar rotational jumps of $\pi/3$ in 2D polydisperse supercooled liquids that possess similar HBOO, as will be discussed in Sec.~\ref{subsec:poly}.

The rotational jumps of the dumbbell probes also affect their bond decorrelation dynamics ($U(t)$), as shown in Fig.~\ref{fig:tcf}(A). In the isotropic liquid phase ($\phi<0.7$), $U(t)$ decays exponentially, satisfying the Gaussian approximation, \emph{i.e.}, $U(t)=\exp[-\langle(\Delta \varphi(t))^2\rangle/ 2]$. This is consistent with the previous observation that dumbbells behave as Brownian rotors in the liquid phase, exhibiting dynamical homogeneity. In the presence of the HBOO ($\phi\geq0.7$), however, $U(t)$ begins to decay in a stretched-exponential fashion, deviating from the Gaussian approximation. $U(t)$ is well fitted by the KWW function (Eq.~\ref{eq:kww}) with the stretching exponent $\beta = 0.85$, $0.88$, and $0.9$ for $\phi = 0.7$, $0.71$, and $0.72$, respectively. This deviation is consistent with the previous results of systems exhibiting infrequent rotational jumps~\cite{StillingerJCP,StarrPRE,ChongPRL}. 

\begin {figure}
\centering\includegraphics [width=3in] {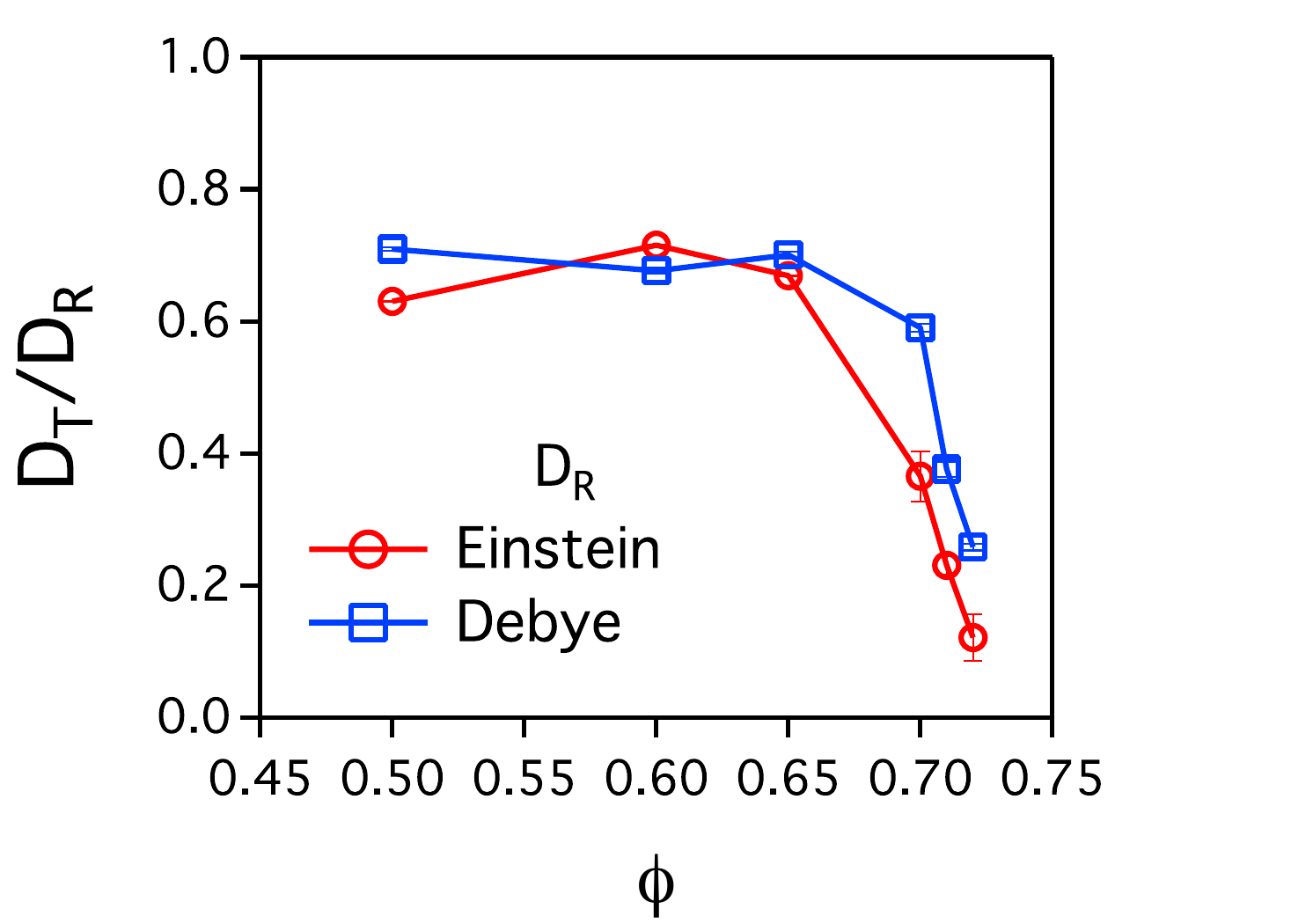}
\caption{Ratio of translational ($D_T$) to rotational ($D_R$) diffusion coefficients of dumbbell probes as a function of area fraction ($\phi$). Red and blue symbols denote $D_T/D_R$ obtained from the Einstein and Debye formalisms, respectively.}
\label{fig:ratio}
\end{figure}

\begin {figure*}
\centering\includegraphics [width=6.0in] {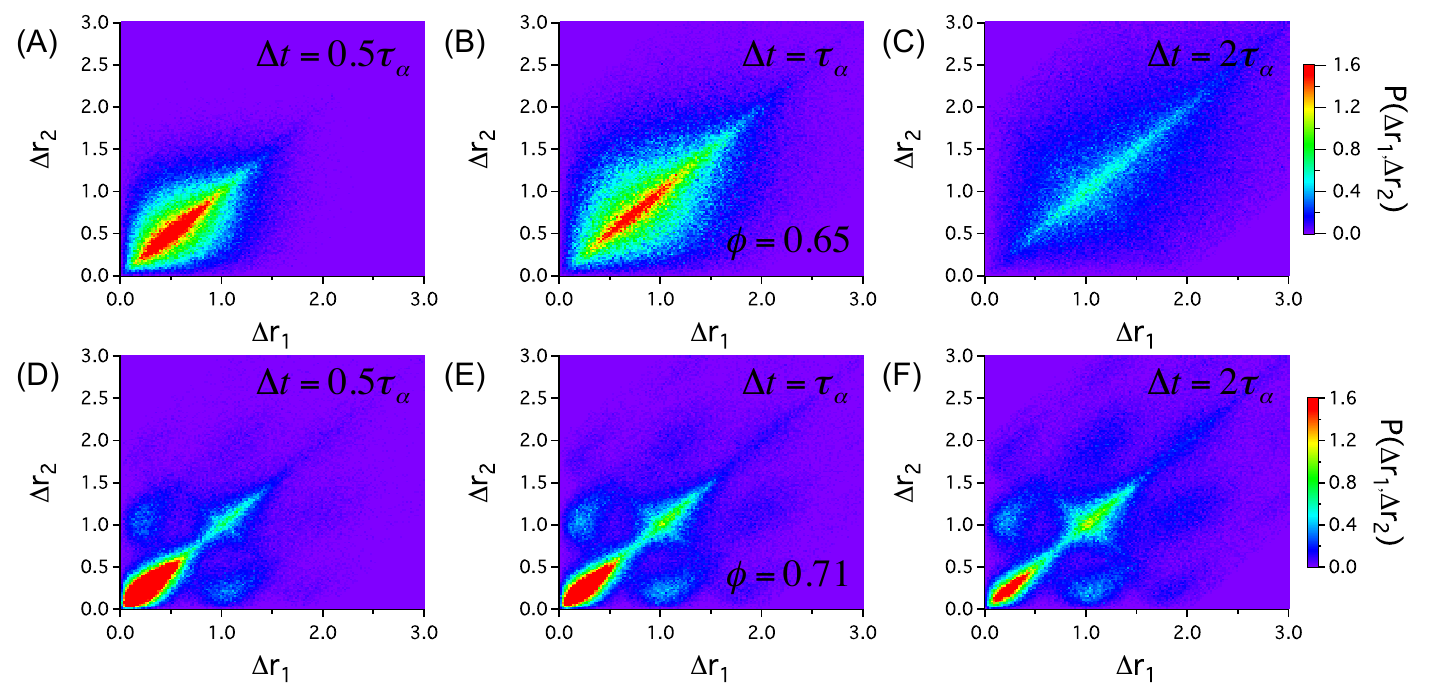}
\caption{Probability distribution functions $P(\Delta r_1(t),\Delta r_2(t))$ of the mobilities of the two constituent particles of dumbbell probes. Upper panels: results in the 2D liquid phase at $\phi = 0.65$ for (A) $\Delta t = 0.5\tau_\alpha$, (B) $\tau_\alpha$, and (C) $2\tau_\alpha$. Lower panels: results in the hexatic phase at $\phi = 0.71$ for (D) $\Delta t = 0.5\tau_\alpha$, (E) $\tau_\alpha$, and (F) $2\tau_\alpha$.}
\label{fig:coupling}
\end{figure*}

Furthermore, the stretched-exponential decay of $U(t)$ indicates that the dumbbell probes experience and report a distribution of dynamic environments in the hard-disc medium. Figure~\ref{fig:tcf}(B) displays a displacement map of discs during a time interval $\Delta t$ (= $t_1 - t_0 = 200\tau$) and illustrates the dumbbell probes at their original positions at $t = t_0$. The colors of the dumbbell probes are assigned according to their maximum rotational displacement $\varphi^{\mathrm{max}}$ during $\Delta t = 200 \tau$. Consistent with previous studies~\cite{vanderMeer2015}, the diffusion of the host discs in the hexatic phase is spatially heterogeneous: fast and slow particles coexist and form distinct domains. The dumbbell probes experience both slow and fast domains. Dumbbells in the fast domains are more likely to rotate with large $\varphi^{\mathrm{max}}$, while those in the slow domains librate with small $\varphi^{\mathrm{max}}$. This heterogeneity, which would individually lead to exponential decays with distinct characteristic times, is averaged over and manifests as a stretched-exponential decay~\cite{EdigerAnnuRev,Mandel2022nat,Mandel2022,Hyeyoung2025} The decay of $U(t)$ for a single dumbbell also follows a stretched-exponential form, since it experiences multiple dynamic environments over time. The microscopic diffusion mechanism of the dumbbell probes will be discussed in Sec.~\ref{subsec:swing}.

Consistent with the presence of dynamic heterogeneities in both translation and rotation, the Debye-Stokes–Einstein relation (DSER) breaks down (Fig.~\ref{fig:ratio}), as evidenced by a decrease in $D_T/D_R$ with increasing $\phi$. We find that the rotational diffusion coefficient $D_R$ of the dumbbell probes is only marginally sensitive to the formalism used: both Einstein and Debye formalisms yield a decreasing trend in $D_T/D_R$ for $\phi\geq0.7$. Previously, some supercooled liquids~\cite{ChongPRL} exhibited discrepancies in $D_R$ depending on the formalism used, which were attributed to a substantial delay in $U(t)$—by more than an order of magnitude—from the Gaussian approximation. In our system, however, the deviation of $U(t)$ from its Gaussian approximation remains relatively small, even at $\phi = 0.72$ (Fig.~\ref{fig:tcf}(A)). This smaller deviation may be due to the limited jump angle of approximately $\pi/3$, in contrast to the larger angle of $\sim\pi$ observed in supercooled liquids.

\subsection{Swing motion of dumbbell probes in the presence of host HBOO}\label{subsec:swing}

At the single-molecule level, the dumbbell probes can exhibit two diffusion modes in the presence of the HBOO: swing and gliding. It was reported that in degenerate crystals of dicolloidal particles~\cite{CohenPRL1}, dislocation occurs through a combination of these two modes. The swing refers to a motion in which one particle of the dumbbell hops to another site, while the other particle remains at the original site. That is, translation and rotation are coupled during a swing motion. In contrast, gliding refers to a motion along the molecular axis, where translation and rotation are completely decoupled at the single-molecule level.

The tendency of dumbbell probes to exhibit higher rotational diffusion in more mobile hard-disc domains (Fig.~\ref{fig:tcf}(B)) suggests that swing motion is the dominant diffusion mechanism of the probes. In other words, their rotation is strongly coupled to the motion of the surrounding discs, giving rise to swing motion. By contrast, gliding should keep $\varphi^{\mathrm{max}}$ small; such decoupling of rotation from the mobility of the hard-disc medium is inconsistent with the observed correlation between $\varphi^{\mathrm{max}}$ and medium mobility. 

\begin {figure*}
\centering\includegraphics [width=6.5in] {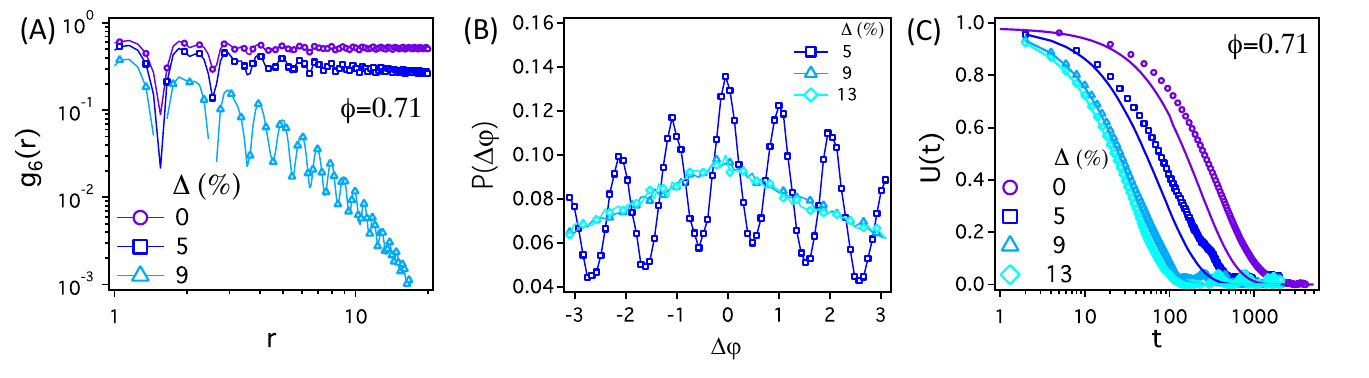}
\caption{Size polydispersity ($\Delta$)–induced 2D melting transition and rotational dynamics of dumbbell probes at $\phi = 0.71$. (A) Bond-orientational correlation function $g_6(r)$. (B) Distribution $P(\Delta \varphi)$ of angular displacements over a time $t^*$ such that $\langle(\Delta\varphi(t^*))^2\rangle=4\pi^2$. (C) Reorientational time correlation function $U(t)$. In panel C, symbols denote simulation results, and solid lines show the exponential decay predicted by the Debye approximation.}
\label{fig:poly}
\end{figure*}

Our additional analysis further supports that the dumbbell probes in the 2D hexatic and solid phases diffuse primarily via swing motion. To distinguish the dominant diffusion modes of dumbbell probes, we compute the joint probability distribution function ($P(\Delta r_1(t), \Delta r_2(t))$) of the mobilities of the two constituent particles of a dumbbell probe, as shown in Fig.~\ref{fig:coupling}. Here, $\Delta r_j(\Delta t) = |\vec{r}_j(t_0+\Delta t)-\vec{r}_j(t_0)|$ with $j \in \{1,2\}$ denoting the particle index of a given dumbbell. Consistent with the results in Fig.~\ref{fig:gstheta}, $P(\Delta r_1(t), \Delta r_2(t))$ differs qualitatively across the freezing transition. In the 2D liquid phase ($\phi = 0.65$, upper panels in Fig.~\ref{fig:coupling}), the intensity of $P(\Delta r_1(t), \Delta r_2(t))$ exhibits a single peak (red region) along the diagonal line with $\Delta r_1(t) \approx \Delta r_2(t)$, indicating that the mobilities of the two constituent particles are highly correlated. As the time interval $\Delta t$ increases from $0.5\tau_\alpha$ to $2\tau_\alpha$, the peak broadens uniformly.

In contrast, in the hexatic phase ($\phi = 0.71$, lower panels), $P(\Delta r_1(t), \Delta r_2(t))$ displays multiple peaks. Notably, off-diagonal peaks near $(\Delta r_1(t), \Delta r_2(t)) = (1\sigma, 0)$ and $(0, 1\sigma)$  emerge in addition to the diagonal peaks. These off-diagonal features reflect the swing motion of a dumbbell probe; for instance, the peak at $(1\sigma, 0)$ corresponds to the case where the first particle hops while the second remains stationary. The peaks along the diagonal line, such as $(\Delta r_1(t), \Delta r_2(t)) = (1\sigma, 1\sigma)$, represent the contributions of a sequence of multiple swing motions and gliding motion~\cite{Oh2016}. Yet, we find no significant anisotropy in the diffusion of the dumbbell probes, supporting the swing motion as a dominant diffusion mechanism. The multiple peaks persist even beyond the structural relaxation time $\tau_\alpha$, clearly revealing the heterogeneous rotation of the dumbbells in the diffusive regime, in agreement with the analysis of $G(\varphi,t)$ (Fig.~\ref{fig:gstheta}).

\subsection{Size polydispersity–induced 2D melting highlights the role of host HBOO in non-Gaussian probe dynamics}\label{subsec:poly}

The size polydispersity ($\Delta$), in addition to packing density, serves as another control parameter for the phase behavior of 2D colloids~\cite{SadrPRL,ItoPRE,Kawasaki:2011p1594,Tanaka:2011p1602,Tanaka2025}. Figure~\ref{fig:poly} demonstrates the reporting capability of dumbbell probes across the melting transition, from the hexatic to the isotropic-liquid phase, induced by increasing $\Delta$ at a fixed $\phi = 0.71$. Figure~\ref{fig:poly}(A) shows the transition in a bond-orientational correlation function ($g_6(r)$) of colloidal discs from a quasi-long-ranged to short-ranged HBOO, indicating a hexatic-to-liquid transition with increasing $\Delta$ from 0 to 0.09. Following the reentrant melting, the dynamics of the hard-disc medium, characterized by short-ranged HBOO, become faster and less heterogeneous. We find that the dumbbell probes effectively report such dynamical changes in 2D polydisperse colloids. As depicted in Fig.~\ref{fig:poly}(B), in case of $\Delta = 0.05$, rotational jump motions by $\pi/3$ lead to a sinuous angle distribution function $P(\Delta \varphi)$, consistent with the oscillatory $G(\varphi,t)$ (Fig.~\ref{fig:gstheta}). Here, $\Delta \varphi = \varphi(t_0+t^*)- \varphi(t_0)$ with $t^*$ being a time when $\langle (\Delta \varphi(t^*))^{2}\rangle = 4\pi^2$. However, when $\Delta \gtrsim 0.09$ with the short-ranged HBOO, $P(\Delta \varphi)$ becomes smooth with no obvious rotational jumps. 

In addition, $U(t)$ of dumbbells (Fig.~\ref{fig:poly}(C)) also captures the reentrant melting transition, with its decay changing from stretched-exponential to exponential behavior. As expected from the results in the monodisperse medium (Fig.~\ref{fig:tcf}(B)), $U(t)$ deviates from the Gaussian approximation in the hexatic phase ($\Delta\leq0.05$) as dumbbells exhibit rotational jumps. In 2D isotropic liquids ($\Delta\geq0.09$), $U(t)$ well obeys the approximation with an exponential decay, as they behave as Brownian rotors. Our results corroborate that the rotational jumps by $\pi/3$ and non-Gaussian nature of the rotational dynamics of dumbbell probes correlate strongly with the HBOO of the 2D colloidal medium.

\section{Summary and Conclusions}\label{sec:conclusion}
In this work, we study the non-Gaussian rotational diffusion of dumbbell probes in 2D hard disc colloids by conducting DMD simulations. In 2D, the HBOO emerges after the freezing transition occurs ($\phi\geq0.7$), which significantly affects both translation and rotation of dumbbell probes. After the HBOO emerges, both $\langle (\Delta r(t))^2 \rangle$ and $\langle (\Delta \varphi (t))^2 \rangle$ of dumbbells start to show subdiffusion at intermediate times. Moreover, $G(\varphi,t)$ becomes oscillatory at $\phi$ = 0.71, whereas it remains Gaussian in the isotropic liquid phase. The peaks of the oscillatory $G(\varphi,t)$ indicate rotational jumps of $\pi/3$, arising from the underlying HBOO in the hexatic and solid phases. Interestingly, these oscillations persist even when $\langle (\Delta \varphi(t))^2 \rangle$ grows linearly with time, i.e., in the diffusive regime. In other words, the rotational dynamics appear Fickian yet remain heterogeneous, analogous to the translational diffusion of 2D hard discs~\cite{vanderMeer2015}. 

The reporting dynamics of the dumbbell probes faithfully reflects the dynamic heterogeneity of the host colloids as evidenced by the stretched $U(t)$, which represents  averaging over several dynamic regions of different mobility in 2D hexatic and solid phases. Although the Gaussian approximation for probe rotation fails in these phases, no discrepancy in determining $D_R$ is observed between the Debye and Einstein formalisms. The ratio $(D_T/D_R)$ exhibits the same trend as a function of $\phi$, regardless of the formalism used~\cite{StillingerJCP,Kawasaki2019,Zendehroud2024}. Such agreement between the two formalisms is partly due to the relatively small jump angle $(\pi/3)$, compared with the $\pi$ jumps observed in supercooled liquids. Despite the observed conventional decoupling, the dumbbell probes also exhibit the coupling between translation and rotation at the single molecule level, as they diffuse primarily via swing motion, rather than gliding, under the HBOO. The heterogeneity of the probe dynamics in the 2D hexatic and solid phases disappears as the host system reenters the liquid phase driven by the size polydispersity $\Delta \gtrsim 0.09$ at $\phi$ = 0.71. This further demonstrates that the probe dynamics well reflects the structural and dynamical features of the host medium.
 
\begin{acknowledgments}
This work was supported by the National Research Foundation of Korea(NRF) grant funded by the Korea government(MSIT) (RS-2024-00338551).  This research was supported by the Regional Innovation System \& Education(RISE) program through the Jeju RISE center, funded by the Ministry of Education(MOE) and the Jeju Special Self-Governing Province, Republic of Korea.(2025-RISE-17-001).
\end{acknowledgments}

 \bibliography{dimer}

\end{document}